\documentclass[fleqn,usenatbib]{mnras}

\usepackage{newtxtext,newtxmath}

\usepackage[T1]{fontenc}

\usepackage{graphicx}	
\usepackage{amsmath}	
\usepackage[normalem]{ulem}

\title[Keplerian motion of a gap-carving planet]{Confirmation and Keplerian motion of the gap-carving protoplanet HD~169142~b}

\author[I. Hammond et al.]{
Iain Hammond,$^{1,2}$\thanks{E-mail: iain.hammond@monash.edu}
Valentin Christiaens,$^{2}$
Daniel J. Price,$^{1}$
Claudia Toci,$^{3}$
Christophe Pinte,$^{1,4}$\newauthor{
Sandrine Juillard$^{2}$
and Himanshi Garg$^{1}$}
\\
$^{1}$ School of Physics \& Astronomy, Monash University, Vic 3800, Australia\\
$^{2}$ Space sciences, Technologies \& Astrophysics Research (STAR) Institute, Universit\'e de Li\`ege, All\'ee du Six Ao\^ut 19c, B-4000 Sart Tilman, Belgium\\
$^{3}$ European Southern Observatory (ESO), Karl-Schwarzschild-Str 2, D-85748 Garching, Germany\\
$^{4}$ Univ. Grenoble Alpes, CNRS, IPAG, F-38000 Grenoble, France\\
}
\date{Accepted XXX. Received YYY; in original form ZZZ}

\defcitealias{Gratton:2019tc}{G19}

\pubyear{2023}

\begin{document}
\label{firstpage}
\pagerange{\pageref{firstpage}--\pageref{lastpage}}
\maketitle

\begin{abstract}
We present the re-detection of a compact source in the face-on protoplanetary disc surrounding HD 169142, using VLT/SPHERE data in \textit{YJH} bands. The source is found at a separation of 0\farcs319 ($\sim$37~au) from the star. Three lines of evidence argue in favour of the signal tracing a protoplanet: (i) it is found in the annular gap separating the two bright rings of the disc, as predicted by theory; (ii) it is moving at the expected Keplerian velocity for an object at $\sim$37~au  in the 2015, 2017 and 2019 datasets; (iii) we also detect a spiral-shaped signal whose morphology is consistent with the expected outer spiral wake triggered by a planet in the gap, based on dedicated hydrodynamical simulations of the system. 
The \textit{YJH} colours 
we extracted for the object are consistent with 
tracing scattered starlight, 
suggesting that the protoplanet is enshrouded in a significant amount of dust, as expected for a circumplanetary disc or envelope surrounding a gap-clearing Jovian-mass protoplanet.


\end{abstract}

\begin{keywords}
protoplanetary discs -- planet-disc interactions -- stars: individual: HD~169142
\end{keywords}



\section{Introduction}\label{sec:intro}

The lack of a statistically significant number of unambiguous protoplanet detections limits our understanding of giant planet formation. 
The protoplanetary disc of HD~169142 is a prime location to search for and expand the sample of directly imaged protoplanets.
HD~169142 is a nearby (114.8 pc, \citealt{Gaia-Collaboration:2022}), 6-Myr old, 1.85M$_{\odot}$ (\citealt{Gratton:2019tc}; hereafter \citetalias{Gratton:2019tc}) Herbig Ae star surrounded by a protoplanetary disc seen almost face-on (\textit{i}=13\degr, PA=5\degr; \citealt{Raman:2006}). 
Observations have revealed two sets of rings at 25 and 60~au delimit a 22-au radius central cavity and an annular gap at 38~au, both depleted in dust and gas \citep{Honda:2012,Quanz:2013te,
Pohl:2017ub,Macias:2017,Fedele:2017wn,Macias:2019, Perez:2019um, Garg:2022}. 
Dedicated hydrodynamical simulations suggest the presence of two 2-4M$_{\rm J}$ planets, in the cavity and gap respectively 
\citep{Toci:2020aa}. 

\begin{table*}
\begin{center}
\caption{Summary of the archival VLT/SPHERE high-contrast imaging observations of HD~169142 used in this work.}
\label{tab:obs}
\begin{tabular}{lccccccccccccc}
\hline
\hline
Date & Strategy & Program & Instrument & Filter & Plate scale$^{\rm (a)}$ & Coronagraph & DIT & 
T$_{\rm int}^{\rm (b)}$ & $<\beta>^{\rm (c)}$ & $\Delta$PA$^{\rm (d)}$ & SNR$_{\rm b}^{\rm (e)}$\\
& &  & &  & [mas px$^{-1}$] & & [s] & 
[min] & [\arcsec] & [\degr] \\
\hline

2015-05-03 & PDI & 095.C-0273(A) & IRDIS & \emph{J} & 12.263$\pm 0.009$  & N\_ALC\_YJ\_S & 32  & 
53.3 & 0.44 & -- & -- \\

2015-06-28 & ASDI & 095.C-0298(C) & IFS & \emph{YJH} & 7.46$\pm 0.02$ & N\_ALC\_YJH\_S & 64  & 
58.7 & 1.00 & 35.0 &4.0\\

2017-04-30 & ASDI & 198.C-0209(G) & IFS & \emph{YJH} & 7.46$\pm 0.02$ & -- & 2  & 
38.4 & 0.62 & 97.8 & 4.3\\

2019-05-19$^{\rm (f)}$ & ASDI & 1100.C-0481(N) & IFS & \emph{YJH} & 7.46$\pm 0.02$ & N\_ALC\_Ks & 96 & 76.8 & 0.59 & 118.2 & 4.5\\

2019-05-19$^{\rm (f)}$ & ADI & 1100.C-0481(N) & IRDIS & \emph{K12} & 12.265$\pm 0.009$ & N\_ALC\_Ks & 96  & 76.8 & 0.59 & 118.2 & --\\
\hline
\end{tabular}
\end{center}
Notes: $^{\rm (a)}$\citet{Maire:2016vq}. $^{\rm (b)}$Total integration time 
after bad frame removal. $^{\rm (c)}$Average seeing at $\lambdaup$~=~500nm. 
$^{\rm (d)}$Parallactic angle variation. $^{\rm (e)}$Signal-to-noise ratio of the protoplanet achieved in the processed image, following the definition in \citet{Mawet:2014uw}. $^{\rm (f)}$New to this work.
\end{table*}

Protoplanet candidate detections were first claimed around HD~169142 based on high-contrast VLT/NACO $L'$-band and MagAO near-IR observations \citep{Reggiani:2014vi,Biller:2014}. 
Polarimetric differential imaging (PDI, \citealt{Kuhn:2001}) observations with the Spectro-Polarimetic High contrast imager for Exo-planets REsearch (SPHERE; \citealt{Beuzit:2008}) have also
revealed a planet candidate in the disc's annular gap at $\sim$36 au separation. 
\citet{Pohl:2017ub} presented $J$-band PDI of HD 169142 at $\lambdaup$\textsubscript{$c$}~=~1.24$\mu$m. Although not discussed at the time, our reduction of this data shows scattered light emission in the annular gap, suggestive of 
a localized collection of dust at a position angle of $\sim$44\degr. \citet{Bertrang:2018us} presented a tentative re-detection in SPHERE/ZIMPOL data at shorter wavelengths although this feature was not mentioned either. 
Based on SPHERE/IFS data, \citet{Ligi:2018ud} suggested that the $L'$ protoplanet candidates from 2014 are tracing filtered signals from the bright inner ring, while also presenting a bright structure at 93 mas interior to the inner ring. 
\citetalias{Gratton:2019tc} claimed the first possible detection of the outer planet in the annular gap using SPHERE/IRDIFS observations (their ``blob D''). However, several observations 
used in their work reached insufficient contrast for a firm re-detection of their candidate, due to a combination of sub-optimal observing conditions, short integration time, and aggressive post-processing. Their candidate was later found to overlap with a non-Keplerian kinematic feature \citep{Garg:2022}. 

In this Letter, we present the re-detection of the compact source in the gap between the two dust rings at three epochs, as well as the first detection of its outer spiral wake, 
by re-reducing archival SPHERE/VLT data with state-of-the-art point-spread-function subtraction techniques. 
We show source movement over the four years between observations consistent with Keplerian orbital motion.

\section{Observations and Data Reduction}\label{sec:observations}

\label{sec:pdi} 
\citet{Pohl:2017ub} first published the 2015 data, employing Polarimetric Differential Imaging (PDI). The \texttt{N\_ALC\_YJ\_S} coronagraph was used to attenuate light from the star, allowing for 32s exposures in $J$-band. Five polarimetric cycles were captured, each consisting of four datacubes in the four Stokes parameters.
We re-reduced this dataset using the IRDIS Data reduction for Accurate Polarimetry pipeline (\textsc{irdap 1.3.4}, \citealt{van-Holstein:2020vk}) with default parameters to produce a Q$_{\phi}$ image. We interpolated the star position using the centre calibration files from the start and end of the observation.

\label{sec:irdifs}

We re-reduced
SPHERE observations of HD~169142 obtained in \texttt{IRDIFS-EXT} mode, corresponding to simultaneous acquisition of Integral Field Spectrograph (IFS, \citealt{Claudi:2008ut}) images in the \textit{YJH}-bands and IRDIS images with the K1 and K2 filters ($\lambda$ $\approx$ 2.11 and 2.25µm, respectively). 
These are summarized in Table~\ref{tab:obs}.
Our selection was motivated by needing long integration times and large bandwidth. The 2015 and 2017 data sets were published in \citet{Ligi:2018ud}, while the 2019 data is new to this work. We only considered the 2019 epoch of SPHERE/IRDIS \emph{K12} data, as it corresponded to the best observing conditions and integration time.

We reduced both IFS and IRDIS data using an in-house pipeline, \textsc{vcal-sphere}\footnote{\url{https://github.com/VChristiaens/vcal_sphere}} \citep{Christiaens:2021}. The pipeline uses the ESO Recipe Execution Tool \textsc{esorex} (v3.13.6) 
for calibration, whereas for pre- and post-processing we used 
the Vortex Image Processing package (\textsc{vip}; \citealt{Gomez-Gonzalez:2017uw}; Christiaens et al. subm.~to JOSS). For IRDIS, our pipeline 
involved: 
(i) flat-fielding, (ii) Principal Component Analysis (PCA, \citealt{Amara:2012um})-based sky subtraction, (iii) bad pixel correction, (iv) centring of the coronagraphic images via satellite spots, (v) centring of the non-coronagraphic unsaturated PSF images using 2D Gaussian fits, (vi) anamorphism correction, (vii) fine centring and bad frame removal using a bright background star, (viii) final PSF creation and (ix) post-processing with median-angular differential imaging \citep[ADI;][]{Marois:2006vd}, PCA-ADI in full-frame and annular PCA-ADI. 
The final $K1$ and $K2$ images are combined to further enhance the SNR of faint circumstellar signals (hereafter referred to as \emph{K12} image). 

Reduction of the IFS data involved additional steps compared to IRDIS: master darks, master detector and instrument flats, sky cubes, spectra positions and wavelength calibration were computed to produce calibrated spectral cubes of 39 channels. \textsc{vip} functions handled pre-processing as for our IRDIS data (see above). Our pipeline 
then leveraged both angular and spectral differential imaging (SDI) using PCA (i.e.~PCA-ASDI) 
for an optimal subtraction of speckles, through the corresponding routines of \textsc{vip}. We used 
annular PCA-ASDI 
in two steps (SDI+ADI) on the 2015 and 2017 datasets, as they were acquired in mediocre observing conditions, and 
in a single step for the 2019 dataset (see details on each PCA-ASDI flavour in \citealt{Christiaens:2019}). These choices led to the best achieved contrasts, respectively. We explored 1–20 principal components.

\begin{figure*}
    \includegraphics[width=\textwidth]{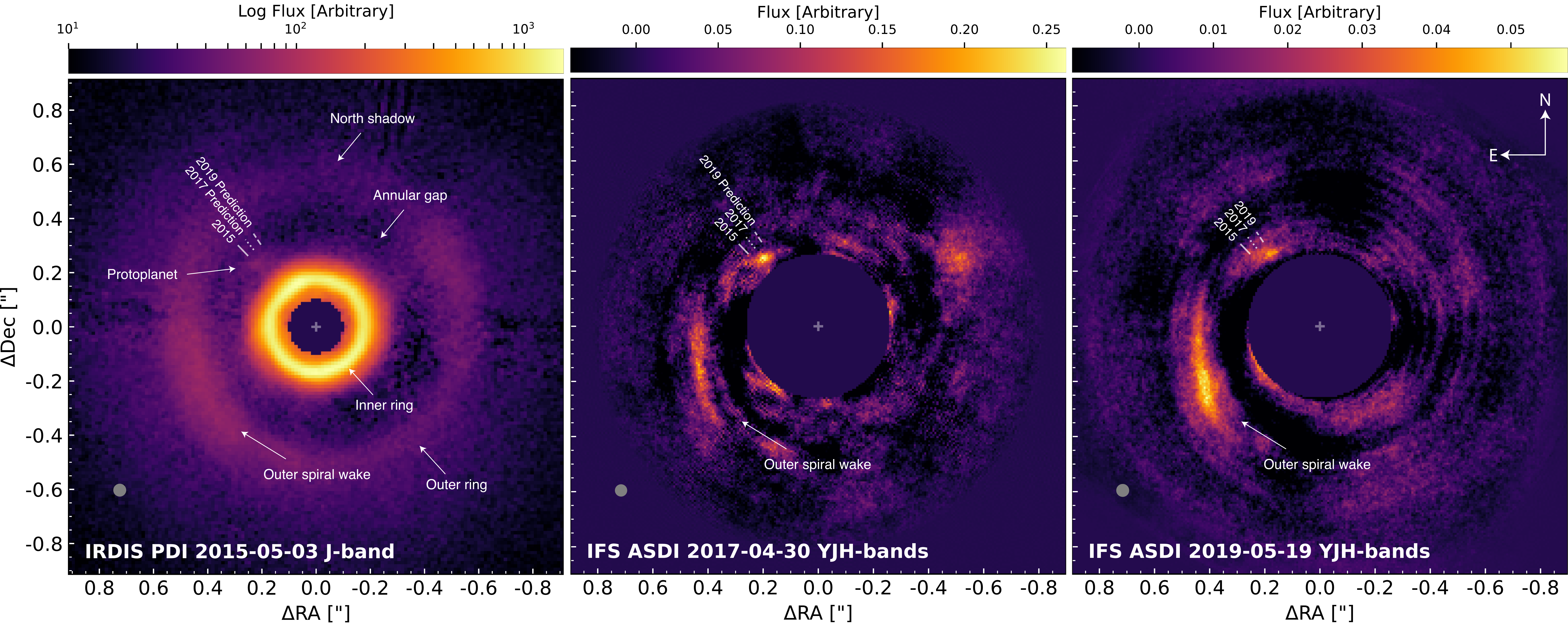}
    \caption{SPHERE near-infrared observations of the disc surrounding HD~169142. \textit{Left:} IRDIS PDI data tracing light scattered off small grains, showcasing a concentration of grains in two rings and a localized collection of dust in the annular gap, suggestive of circumplanetary material. The inner working angle of the coronagraph has been masked. \textit{Middle and right:} Detection of the protoplanet and accompanying outer spiral wake with SPHERE/IFS. The bright inner disc has been masked. In each image, the grey patch represents the FWHM of the non-coronagraphic images obtained during the observing sequence.\label{fig:obs}}
\end{figure*}

\section{Results}\label{sec:results}

\subsection{Compact source in the annular gap}\label{sec:compact_source}

Figure~\ref{fig:obs} shows our new reduction of the PDI dataset (left) alongside our re-detection of both the compact source in the gap and the first detection of spiral-shaped signal in its wake in the 2017 and 2019 IFS datasets. The ASDI reduction of the 2015 IFS dataset, of lower quality, is provided in Appendix~\ref{sec:appendixa}. The source is robust to the number of principal components (1-20) and has an optimal SNR of $\sim$4.0, $\sim$4.3 and $\sim$4.5 in the 2015, 2017 and 2019 IFS data sets, respectively, 
after considering small number statistics \citep{Mawet:2014uw}. We consider these values as lower limits for the 2017 and 2019 data given the presence of a negative ADI-induced spiral signature (associated with a true disc feature also seen in PDI images; Sec.~\ref{sec:disc_features}) overlapping with noise apertures. 
Given that the signal is located within the cleared annular gap, it does not trace filtered ring signal. 

We measured the position of the compact source to be $r$~=~0\farcs319$\pm$0\farcs013 and PA~=~33.9$\pm$2.3\degr~in the 2019 IFS dataset. We used the negative fake companion technique \citep[NEGFC;][]{Lagrange:2010} combined with PCA-ADI applied to each spectral channel, before median-combining the residuals along the wavelength axis. We first used a Nelder-Mead simplex algorithm to obtain a guess on the (negative) planet flux and location which minimized pixel intensities in the vicinity of the planet. 
We then fed this estimate to a Markov chain Monte Carlo 
algorithm 
to determine the posterior probability distribution of the three parameters after 2,500 steps (30\% burn-in). A new expression for the log-probability function was used to account for residual speckle uncertainty on the parameters (\citealt{Christiaens:2021}). As HD~169142 rotates fast on the sky when viewed from Cerro Paranal, we also considered a 1.2\degr~uncertainty associated to the average smearing over the sequence in 96-s long exposures.

For the PDI data and remaining IFS datasets (where the protoplanet could only be imaged through ASDI) 
we performed a 2D Gaussian fit to the source. We derived 
$r$~=~0\farcs321$\pm$0.010 and PA~=~43.7$\pm$1.7\degr in the PDI image. For the IFS datasets, we derived $r$~=~0\farcs0.319$\pm$0.011 and PA~=~44.1$\pm$2.3\degr~for 2015 and $r$~=~0\farcs313$\pm$0.006 and 39.8$\pm$2.3\degr~for 2017 (i.e. in agreement with the values reported in \citetalias{Gratton:2019tc}). For all datasets, the final quoted uncertainties were found by summing the error on our fits with all sources of systematic error in quadrature, including those associated to star centring ($\sim$5~mas) and to the true north of IFS and IRDIS \citep{Maire:2016vq}. 

Considering a distance of 114.8$\pm$0.4~pc, disc PA=5\degr~and inclination \textit{i}=13\degr, the deprojected physical separation of the source in our IFS data is 37.2$\pm$1.5~au. 
Fig.~\ref{fig:keplerian} shows that our four PA measurements 
are consistent with the predicted Keplerian motion for a planet at 37~au orbiting a 1.85M$_{\odot}$ star. 



\subsection{Disc features} \label{sec:disc_features}
Rings are most obvious in the Q$_{\phi}$ image (see left panel of Fig.~\ref{fig:obs} where we have labelled key disc features). We recover the compact signal at the same location 
as presented in \citet{Pohl:2017ub} in the region with excess scattering between the two dust rings (left panel). Brightness variations in azimuth can be seen on the outer ring: there is an emission deficit to the North and overbrightness to the east. 
Our ASDI images obtained in the best observing conditions (Fig.~\ref{fig:obs}) reveal a 
spiral arm to the east. 
The emission coincides with 
the overbrightness in the outer ring seen in scattered light (left panel), suggesting an unresolved superposition of the spiral arm and the outer dust ring. 
The overall signal morphology in our images suggests emission from a protoplanet 
connected to a trailing spiral arm to the east, which would imply clockwise rotation.

\subsection{Spectrum of the compact source}\label{sec:astrometry}

Given the faintness of the compact source, we created three sub-cubes from a weighted average along the wavelength axis of the 4D IFS cube, following the IRDIS \textit{Y}, \textit{J} and \textit{H}2 filter transmission curves, respectively. Fig.~\ref{fig:annuli} shows that the compact source is re-detected separately in each filter. We did not re-detect any compact signal in the IRDIS \emph{K12} images, however the spiral was recovered. Each cube was post-processed using PCA-ADI 
in a single annulus encompassing the same radial separation as the protoplanet, using the same rotation threshold and annulus width as in Sec.~\ref{sec:irdifs}. We set the number of Principal Components to maximise the SNR of the protoplanet using the definition in \citet{Mawet:2014uw}.

Fig.~\ref{fig:spectrum} shows the contrast spectrum for the source.
We measured the flux in each band using the NEGFC technique as in Sec.~\ref{sec:compact_source}. We derive a contrast with respect to the star of 
1.49$\times$10$^{-5}$$\pm$5.95$\times$10$^{-6}$, 1.72$\times$10$^{-5}$$\pm$6.16$\times$10$^{-6}$ and 1.93$\times$10$^{-5}$$\pm$1.17$\times$10$^{-5}$ for the 
\textit{Y}, \textit{J} and \textit{H2} bands, respectively. Non-detection in the K12 band gives a 3-$\sigma$ upper contrast limit of 5.1$\times$10$^{-5}$.


\section{Discussion}\label{sec:discussion}
\begin{figure}
    \includegraphics[width=\columnwidth]{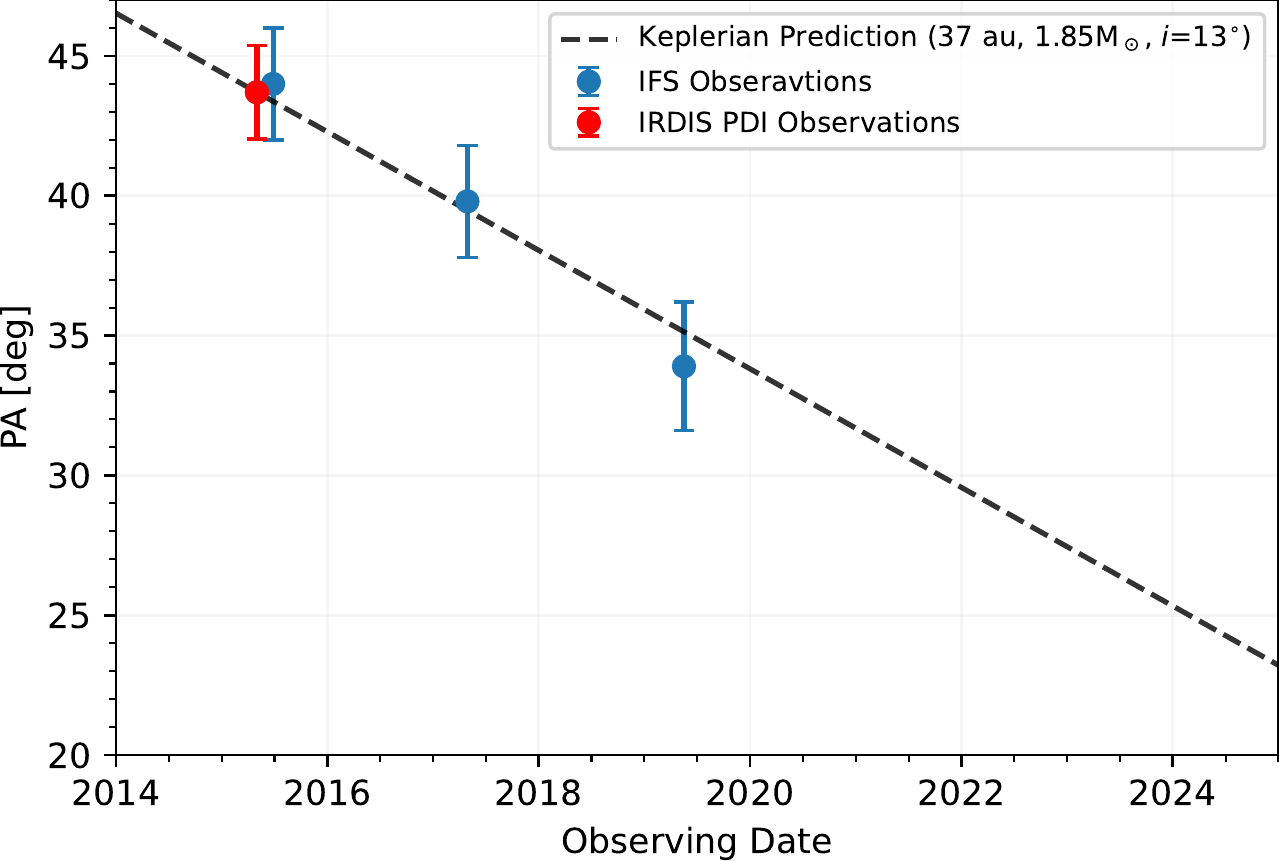}
    \caption{Measured position angle of the protoplanet in the IFS and PDI datasets (blue and red error bars, respectively), compared to expected Keplerian rotation around a 1.85M$_\odot$ star for an object at 37~au on an orbit inclined by 13$\degr$ to the plane of the sky (dashed line).\label{fig:keplerian}}
\end{figure}

The motion of the source matches the expectations for a protoplanet on a circular orbit at a radius of 37~au around a 1.85 M$_\odot$ star (Fig.~\ref{fig:keplerian}). 
Uncertainties (for which we included systematic errors) cannot account for such displacement over four years. Clockwise rotation is consistent with previous studies of the motion of disc substructures 
using multi-epoch imaging (e.g. \citealt{Ligi:2018ud}, \citetalias{Gratton:2019tc}). The compact source coincides with a localized kinematic excess with respect to Keplerian rotation in the gap \citep{Garg:2022}, suggesting an embedded planet.

The depleted annular gap has been explained by the presence of a (super-)Jupiter 
mass planet \citep{Fedele:2017wn,Bae:2018,Lodato:2019,Toci:2020aa}, where 6 Myrs have provided time for a planet to deplete the gap of dust grains. ALMA observations at 1.3mm \citep{Perez:2019um} suggest that these grains have been most strongly filtered from the annular gap as expected for planet-induced clearing \citep{Rice:2006aa}, whereas smaller grains remain coupled to the gas close to the protoplanet as seen in Fig.~\ref{fig:obs} (left). Each tracer has a different radial extent (i.e. smaller gap in sub-mm, larger in mm) which is expected for different grain sizes in the case of clearing by a planet \citep{Pinilla:2012}. As the gap is depleted in gas (but not entirely) we can rule out dead zone dust filtration or photoevaporation. Furthermore, the rings do not correspond to snowlines of gas volatiles \citep{Macias:2019}. 

The presence of a planet also accounts for the spiral arm observed to the east,
as suggested by hydrodynamical simulations and analytical wake models 
\citep{Toci:2020aa,Poblete:2022,Garg:2022}. 
Thermal emission from the edge of the outer 
ring can be rejected as (\textit{i}) the spiral has a non-zero pitch angle (not consistent with a ring) and (\textit{ii}) is limited in azimuthal extent. Such a feature does not appear elsewhere in the disc. We do not resolve the inner wake, which is 
likely overlapping with the bright inner ring, although the 2011 $H$-band Subaru/HiCIAO image of the system shows a tentative extended signal in the annular gap at the expected location of the inner spiral wake (towards North and North east; \citealt{Momose:2015}).
We note a reduction in surface brightness of the outer ring towards PA $\sim$ -15\degr--30\degr~(also seen in \textit{R'}, \textit{I'}, \textit{H} and \textit{J}-band irrespective of instrument, \citealt{Momose:2015,Monnier:2017,Bertrang:2018us,Tschudi:2021,Rich:2022}), which could be caused by shadowing from either the inner ring or the expected inner spiral wake.

\begin{figure}
\centering
   \includegraphics[width=\linewidth]{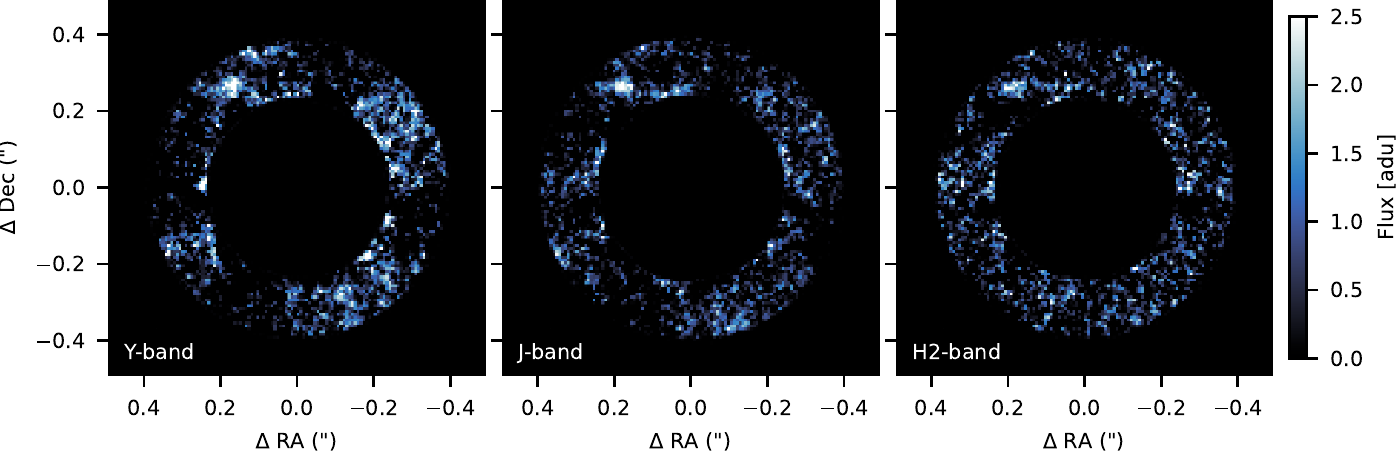}
   \includegraphics[width=\linewidth]{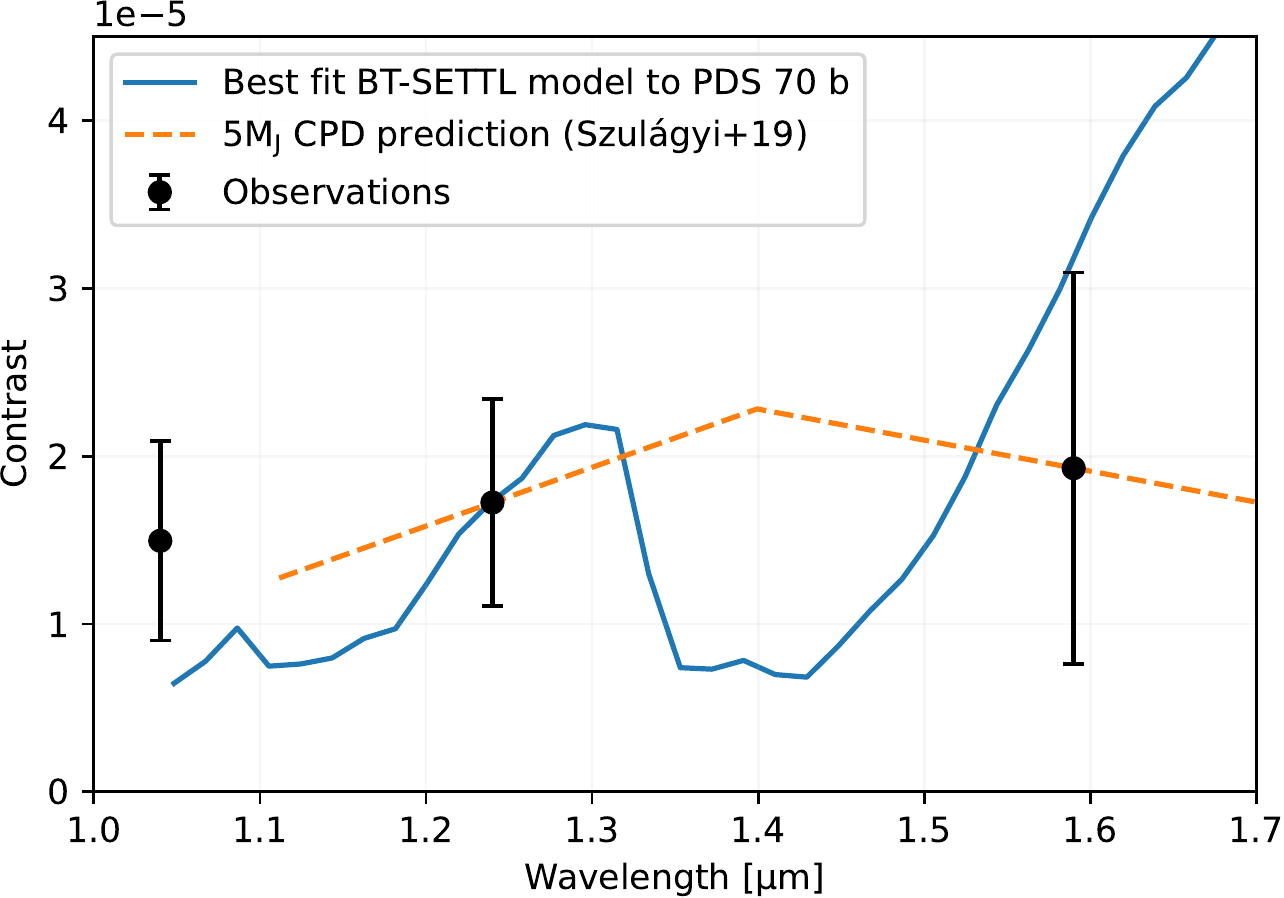}
    \caption{(\emph{top}) Detection of the protoplanet in individual  \textit{Y}, \textit{J} and \textit{H}2-band images using annular PCA-ADI applied to the 2019 IFS dataset. 
    \label{fig:annuli} (\emph{bottom}) Corresponding contrast of the planet with respect to the star (black error bars), 
    compared to the expected contrast spectrum for the best-fit BT-SETTL model of PDS 70 b \citep[][]{Wang:2021b} 
    (blue curve) 
    %
    and for a circumplanetary disc model presented in 
    \citet[][dashed curve]{Szulagyi:2019vo}.
    \label{fig:spectrum}}
\end{figure}

Fig.~\ref{fig:spectrum} compares our contrast spectrum of HD 169142 b with the expected contrast spectrum for PDS 70 b and for circumplanetary disc (CPD) models presented in \citet{Szulagyi:2019vo}, obtained considering the SED of the star presented in \citet{Wagner:2015}. For PDS~70~b, we considered the best-fit BT-SETTL model inferred in \citet{Wang:2021b}. We used \textsc{special} 
\citep{Christiaens2022} to apply the best-fit value found for extinction ($A_{\rm V}$=5.4) and resample the model to the IFS spectral resolution.
The non-detection of the protoplanet in the 2019 IRDIS K12 images 
suggests a $J-K12$ colour lower than 2.1$\pm$0.4 mag. This implies that the colour of HD 169142 b is likely not as red as PDS~70~b 
\citep[$J-K12 \approx 2.2$;][]{Wang:2021b}. The contrast spectrum is consistent with NIR SED predictions for CPD models of either 1 or 5 $M_{\rm J}$ \citep[][]{Szulagyi:2019vo}. 
The spectrum of forming giant planets and their CPD is expected to be dominated by scattered starlight if the planet is embedded, making them potentially detectable in $Q_{\phi}$ polarized maps for favourable (e.g. face-on) geometries \citep[][]{Szulagyi:2019vo,Szulagyi:2021uo}. Our observations suggest this is the case and the protoplanet spectrum likely peaks far into the infrared due to extinction by the circumplanetary disc or envelope. 
Such detection suggests that leveraging polarized light off a planet's circumplanetary dust may be a powerful way to reveal protoplanets in near face-on discs. 

\begin{figure}      \includegraphics[width=\linewidth]{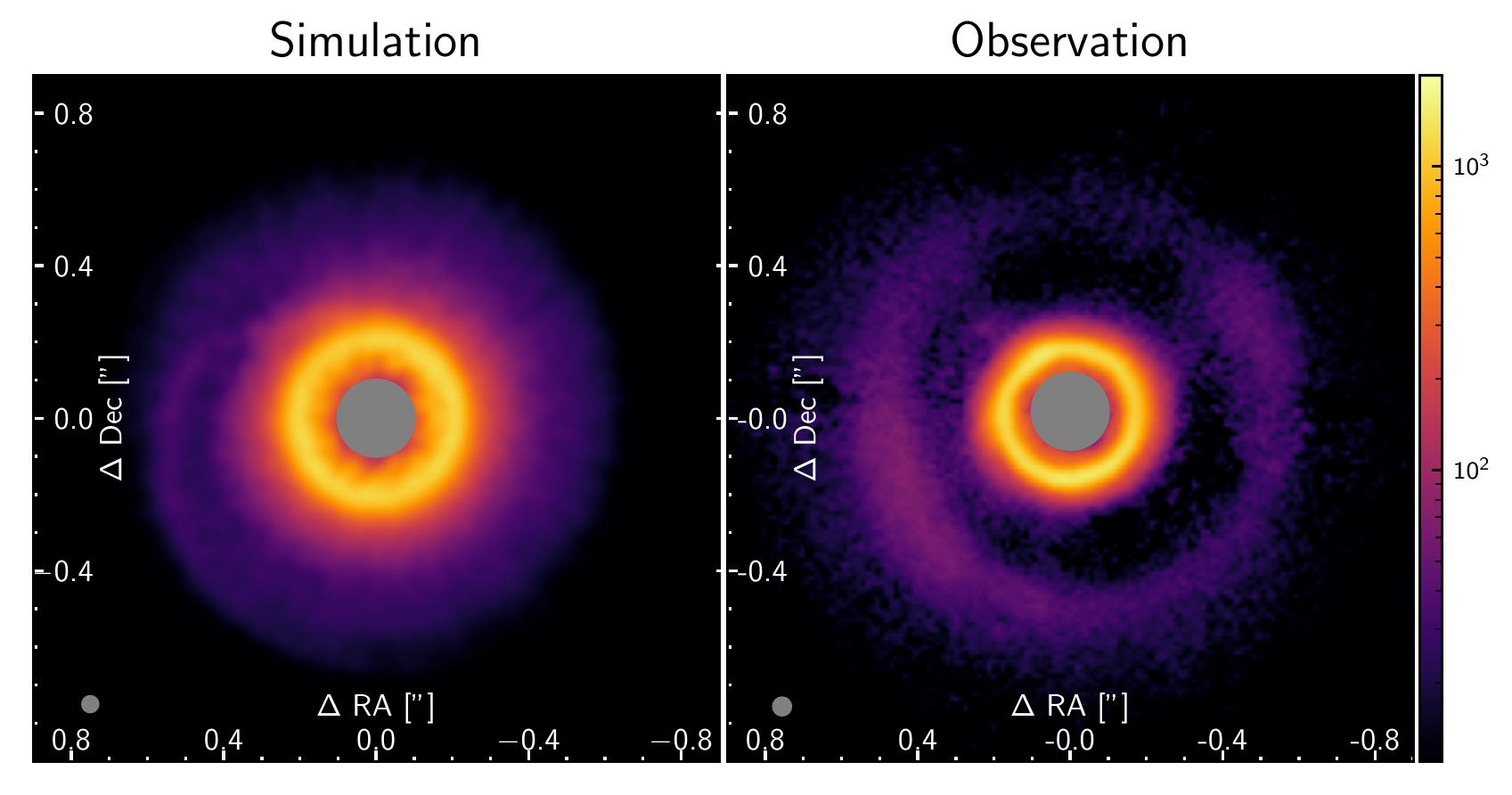}
    \caption{Comparison between the observations and the prediction from the \citet{Toci:2020aa} models.\label{fig:comparison}}
\end{figure}

To show that our observations can be qualitatively explained in the protoplanet scenario, we compare the observed 2015 $Q_{\phi}$ image 
with a synthetic Q$_\phi$ image at 1.25 $\mu$m (Fig.~\ref{fig:comparison}) obtained by processing a snapshot ($\sim$400 orbits, 0.13 Myr of evolution) of the 3D hydrodynamical model presented in \citet{Toci:2020aa} with radiative transfer code \textsc{mcfost} \citep{Pinte:2006uj}. We refer the reader to \citet{Toci:2020aa} for details on the setup of the \textsc{phantom} \citep{Price:2018tl} model. The comparison shows how the presence of a planet located at the detected position of our compact source can qualitatively explain both the annular gap and the spiral wake. The model includes two giant planets (M$\sim 2.5$ M$_{\rm J}$), one located in the inner cavity, at 17 au, and one located in between the two rings, 
at a PA of 44$^\circ$ to match the position of the observed compact source. 
While some emission due to the presence of small dust grains close to the planet can be seen in the model, one should not expect to fully recover the emission as 
circumplanetary material was excluded from the simulation. 
The comparison reveals some limitations of the model: the inner ring is slightly wider, implying that the inner planet should be located 1-2 au inward, and the darkness of the gap is not matched, suggesting either a larger disc aspect ratio or a flared disc (for a discussion, see \citealt{Macias:2019}). We did not attempt to fine-tune these parameters, nor the planet masses and number of orbits, as fresh modelling of this source is beyond the scope of this work.



Observations with longer integration times are required to extract a higher signal-to-noise NIR spectrum of the protoplanet, and follow-up at mid-IR wavelengths may confirm the expected thermal excess of the CPD \citep{Zhu:2015vl,Szulagyi:2019vo}. Deep ALMA sub-mm continuum images may re-detect the CPD, although the outer ring pressure bump may significantly limit the amount of large grains fed to the CPD. Given the amount of small dust surrounding the protoplanet, long-wavelength hydrogen recombination lines (e.g. Br-$\alpha$) may be needed to confirm accretion onto the protoplanet. 
If most protoplanets 
are enshrouded in large amounts of dust as evidenced here, 
extinction may account for the lack of H$\alpha$ detections in discs with strong planet signposts \citep[e.g.][]{Zurlo:2020vi}, except for the most favourable cases corresponding to deep and wide (e.g.~multi-planet carved) gaps. 

\section{Conclusions}

We confirm the protoplanet HD~169142~b first proposed in \citetalias{Gratton:2019tc}, inside an annular gap at 0\farcs319 using angular and spectral differential imaging.
The $10.2\pm2.8{\degr}$ shift in position angle between 2015 and 2019 with intermediate separation in 2017 is consistent with Keplerian motion of the planet. The clockwise motion is consistent with a trailing spiral arm observed to the east. Hydrodynamical simulations from \citet{Toci:2020aa} qualitatively reproduce the observed spiral wake associated with the planet. The contrast spectrum we extracted for the protoplanet suggests starlight scattered off circumplanetary dust, consistent with the 
detection in polarized intensity and expectations from circumplanetary disc models in \citet{Szulagyi:2019vo}.


\section*{Acknowledgements}
We thank Myriam Benisty and Roxanne Ligi for useful discussions.
IH acknowledges a Research Training Program scholarship from the Australian government. IH, DP, CP \& HG acknowledge Australian Research Council funding via DP180104235. VC acknowledges a postdoctoral fellowship from the Belgian F.R.S.-FNRS. CT acknowledges the European Union’s Horizon 2020 research and innovation programme, Marie Sklodowska Curie grant agreement No 823823 (DUSTBUSTERS RISE project). We made use of the Multi-modal Australian ScienceS Imaging and Visualisation Environment (MASSIVE) (\url{www.massive.org.au}) and data from the European Space Agency (ESA) mission \textit{Gaia}.

\section*{Data Availability}
Data will be made available on the Strasbourg astronomical Data Center (CDS) and through Monash University’s \textit{Bridges} research repository via \href{https://doi.org/10.26180/22146428}{doi:10.26180/22146428}.



\bibliographystyle{mnras}
\bibliography{references} 




\appendix

\section{2015 and 2017 SPHERE/IFS data}\label{sec:appendixa}

\begin{figure}\label{fig:year_comparison}\centering
    \includegraphics[width=\columnwidth]{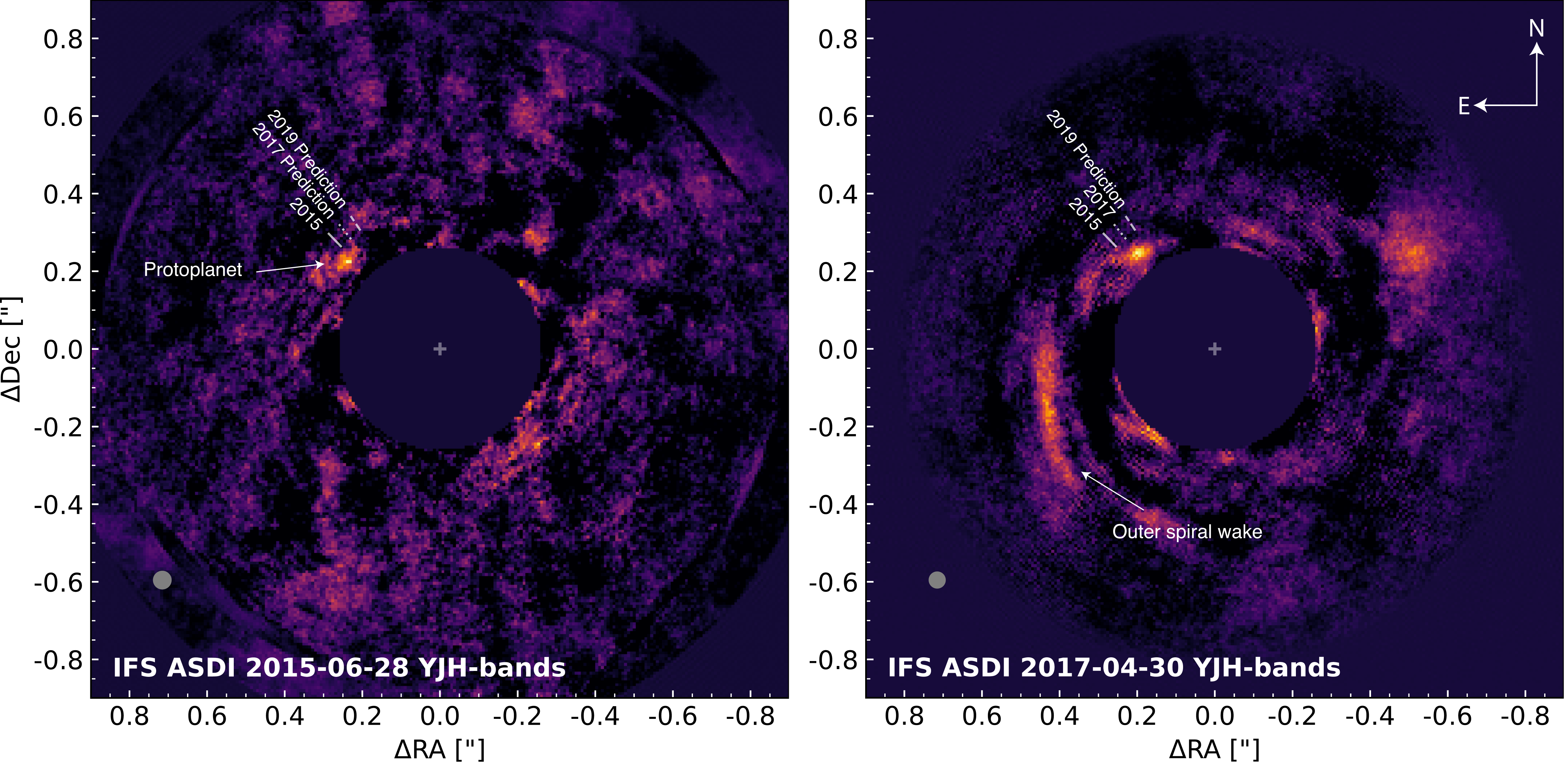}
    \caption{ASDI reductions for the 2015 and 2017 IFS data showcasing motion of the protoplanet. 
    Solid, dotted and dashed ticks indicate the position of the protoplanet in the 2015, 2017 and 2019 IFS data sets, respectively. 
    }
    \label{fig:app}
\end{figure}

Fig.~\ref{fig:app} shows our reduction of 
the 28 June 2015 and 30 April 2017 SPHERE/IFS datasets, 
where we detect the compact emission from the protoplanet at a SNR$\sim$4.0 and 4.3, respectively. The outer spiral wake is recovered in the 2017 dataset (Fig.~\ref{fig:obs}).


\bsp	
\label{lastpage}
\end{document}